\documentclass[useAMS,usenatbib]{mn2e}

\usepackage{epsfig}
\usepackage{myaasmacros}
\usepackage{enumerate}
\usepackage{amsmath}
\usepackage{amsfonts}

\usepackage{comment}

\title[Properties and environmental histories of isolated galaxies] 
{Isolated galaxies in hierarchical galaxy formation models - present-day 
  properties and environmental histories}  
\author[Hirschmann et al.]{Michaela Hirschmann$^{1}$\thanks{E-mail:
mhirsch@oats.inaf.it}, Gabriella De Lucia$^{1}$, Angela Iovino$^{2}$, Olga
  Cucciati$^{1,3}$\\ 
$^{1}$INAF - Osservatorio Astronomico di Trieste, via G.B. Tiepolo 11,
34143 Trieste, Italy\\
$^{2}$INAF - Osservatorio Astronomico di Brera, via Brera 28, 20159
Milano, Italy\\
$^{3}$INAF - Osservatorio Astronomico di Bologna, via Ranzani 1, 40127
Bologna, Italy
}
\begin{document}

\date{Accepted ???. Received ??? in original form ???}

\pagerange{\pageref{firstpage}--\pageref{lastpage}} \pubyear{2002}

\maketitle

\label{firstpage}

\begin{abstract}
In this study, we have carried out a detailed, statistical analysis of
isolated model galaxies, taking advantage of publicly
available hierarchical galaxy formation models. To select isolated
galaxies, we employ 2D methods widely used in the observational
literature, as well as a more stringent 3D isolation criterion that
uses the full 3D-real space information. In qualitative agreement with
observational results, isolated model galaxies have larger fractions
of late-type, star forming galaxies with respect to randomly selected
samples of galaxies with the same mass distribution. We also find that
the  samples of isolated model galaxies typically contain a fraction of
less than $15$ per cent of satellite galaxies, that reside at the
outskirts of their parent  haloes where the galaxy number density is
low. Projection effects cause a contamination of 2D samples of about
$18$ per cent, while we estimate a typical completeness of $65$ per
cent. Our model isolated samples also include a very small (few per
cent) fraction of bulge dominated galaxies ($B/T>0.8$) whose bulges
have been built mainly by minor mergers. Our study demonstrates that
about 65-70 per cent of 2D isolated galaxies that are classified as
isolated at $z=0$ have indeed been completely isolated since $z=1$ and
only 7~per cent have had more than 3 neighbours within a comoving
radius of  $1$~Mpc. Irrespectively of the isolation criteria, roughly
45 per cent of isolated galaxies have experienced \textit{at least}
one merger event in the past (most of the mergers are minor, with mass
ratios between 1:4 and 1:10). The latter point validates the
approximation that isolated galaxies have been mainly influenced by
internal processes.  
\end{abstract}


%*******************************************************************************
%*******************************************************************************
\section{Introduction}\label{intro}
%*******************************************************************************
%*******************************************************************************

It has been known for a long time that galaxy properties depend
strongly on the environment in which they are located. E.g. it has
been shown that red, early-type galaxies are preferentially residing
in high-density regions while blue, late-type spirals represent the
main contribution to galaxy populations in low-density environments
\citep{Oemler74,Dressler80}. In recent years, the completion of large
spectroscopic and photometric surveys has given new impetus to
studies that are trying to assess the role of the environment in galaxy
formation
\citep[e.g.][]{Kauffmann04,Balogh04,Cucciati06,Cooper06}. There
remains, however, an open-issue that as to what extent
the properties of galaxies are determined by external physical
processes that come into play only after galaxies become part of a 
group or of a cluster (`nurture'), or are driven by internal physical
processes (e.g. star formation, AGN feedback - `nature'). Ideally, the
issue should be addressed by using a comparative approach that includes a
sample of galaxies whose physical properties are largely the result of
internal physical processes only. By comparing these galaxies to those
residing in different environments, it should be possible to draw
conclusions on the role played by nurture-induced processes. In this
respect, `isolated galaxies' have long been considered ideal
candidates for a reference sub-sample of galaxies that have not
experienced nurture-related processes during their lifetime.

Many observational attempts have been made to identify isolated
galaxies by using different criteria. In an early study by 
\citet{Kara73}, a catalogue of 1050 isolated galaxies were
  obtained by visual inspection of photometric plates (Catalogue of
Isolated Galaxies: CIG). The isolation criterion used was the
following: a galaxy was classified as isolated if there was no other
galaxy with similar size within 20 times its diameter. The CIG
catalogue is large enough to allow a statistical comparison between
isolated galaxies and galaxies from denser environments
(\citealp{Adams80, Haynes80, Sulentic89, Young86, Sauty03}). In
addition, the sample has a very well defined selection function, and a
completeness of 80-90 per cent. Other early studies of isolated
galaxies are based on only a few tens or a few hundreds of
  galaxies (e.g. \citealp{Huchra77, Vettolani86, Marquez99,
  Marquez00, Colbert01, Pisano02, Varela04}). By comparing isolated
spiral galaxies with spirals at different densities, \citet{Varela04}
showed that isolated spirals tend to have symmetric morphologies and
to be bluer, smaller and less luminous than their non-isolated
counterparts. However, this interesting result is based on a sample
containing only 203 lenticular and spiral isolated galaxies. In
addition, there is no accurate analysis of the completeness of the
sample. 

A similar study using the CIG catalogue is not easy: low-resolution
and the non-linearity of the data (from the Palomar Sky Survey, POSS)
represent major drawbacks for the identification of spiral
galaxies. Galaxy bulges appear larger on low resolution images, and
compact spirals can be misclassified easily as ellipticals or
lenticular galaxies. The CIG catalog was later refined resulting into
the AMIGA (\textbf{A}nalysis of the interstellar \textbf{M}edium of
\textbf{I}solated \textbf{GA}laxies) sample. Work based on this sample
has shown that isolated galaxies have physical properties that differ
systematically from those of field galaxies. In particular, AMIGA
early-type galaxies are usually fainter than late-types, and most
spirals in this sample host pseudo-bulges rather than classical bulges
(\citealp{Verdes05, Sulentic06, Durbala08}). Moreover, the AMIGA
galaxies do not show strong signatures of morphological interactions
(\citealp{Sulentic06}). In a recent study, \citet{Sabater12} show that
a significant fraction of (AMIGA) isolated galaxies can be identified
as optical AGNs, and that there is no difference in the prevalence of
AGNs between isolated galaxies and galaxies in denser
environment. This indicates that major interactions are not a
necessary condition for triggering optical AGNs.

Unfortunately, the term `isolated' has been often used for samples
identified using different criteria, and theoretical studies devoted
to clarify the meaning and interpretation of the adopted definitions
are quite scarce. It is not clear if there is a particular definition
that is better than others, and how results obtained using different
algorithms compare to each other. The issue is largely semantic, but
results from numerical simulations and semi-analytic models can help in
evaluating the performance of different selection algorithms, and support
interpretation of observational results. Recent attempts in this
  direction have been made. For example, a recent study by
  \citet{Muldrew12} uses a mock galaxy catalogue of the nearby
  Universe to compare different density estimators that are commonly
  used in observational studies. In addition, from the theoretical
point of view, a very recent study by \citet{Martig12} analyses a set
of 33 cosmological simulations of the evolution of Milky Way-like
galaxies residing in low density environments. They find that the
progenitors of their sample of simulated `isolated' galaxies span a
wide range of morphological types at z=1, with the most disk-dominated
galaxies having experienced an extremely quiet mass assembly. In an
earlier study, \citet{Niemi10} investigated the properties and the
evolution of isolated field ellipticals using galaxy catalogues based
on the Millennium Simulation. They find that roughly half of these
isolated ellipticals have undergone one major merger event, and that
almost all of them have experienced some merging activity during their
lifetime. 

A detailed investigation, however, based on different selection criteria
mimicking the observational selection, has not been performed yet. In this
work, we present a statistical study of isolated galaxies from a hierarchical
galaxy formation model taking advantage of publicly available galaxy catalogues
based on the Millennium Simulation (\citealp{DeLucia07}). We analyse the
present-day physical properties of model isolated galaxies, as well as their
environmental history. We also consider different selection criteria, comparing
methods widely used in observational studies (e.g. that used to define the
AMIGA sample) with a more stringent 3D isolated criterion that takes advantage
of the full 3D-real space information available in simulations in order to
better understand possible inaccuracies due to projection effects in
observational studies.

The layout of the paper is as follows.  In section \ref{AMIGA}, we
briefly introduce the AMIGA sample and the isolation criteria used in
\citet{Verley07b}, which constitutes the basis for this study. In
section \ref{theory}, we shortly describe the semi-analytic model we
use and how we select model isolated galaxies. Section \ref{galprop}
gives an overview of present-day properties of isolated model
galaxies both for a 2D-observational selection criterion (subsection
\ref{galprop_obs}), and for an alternative criterion based on the full
3D real-space information available from the simulation (subsection 
\ref{galprop_3D}). Section \ref{EnvHist} focuses on the
`environmental history of isolated galaxies', which we quantify by
studying the parent halo mass of the progenitors of isolated galaxies,
as well as their merger activity. Finally, in Section
\ref{discussion}, we summarise and discuss our results, and give our
conclusions.

%*******************************************************************************
%*******************************************************************************
\section{Quantification of isolated galaxies in the AMIGA sample}\label{AMIGA}
%*******************************************************************************
%*******************************************************************************

The AMIGA catalogue is a refinement of the earlier CIG catalogue of 
\citet{Kara73}. The AMIGA sample consists of 950 galaxies classified
as isolated and selected as described in the previous section. The
galaxies included in the sample have apparent magnitudes below
$m_{bj}<15.7$ and recessional velocities larger than $1500\
\mathrm{km/s}$ (\citealp{Verdes05, Verley07b}). For each AMIGA
galaxy, the degree of isolation was estimated considering neighbour
galaxies with apparent magnitudes down to $m_{bj}= 17.5$ and with
recessional velocities below $10,000\ \mathrm{km/s}$ (i.e. the
neighbour-sample is `deeper', and contains more galaxies than the
isolated galaxy sample, see \citealp{Verley07b}). 

Two criteria for defining isolated galaxies were considered: the first
one is based on the neighbour count, and the second one is based on
the tidal strength experienced by each galaxy. For the first
criterion, a local number density ($\Sigma_{5NN}$) is computed by measuring the
distance to the fifth nearest neighbour:
\begin{equation}\label{one}
\Sigma_{5NN} =  \frac{k-1}{4\pi r_k^3/3} ,\
\mathrm{with}\ k=5.
\end{equation}
Here, $r_{5}$ is the distance to the fifth nearest neighbour, and is
in units of arcminutes. When computing the distance from a galaxy $i$
to its surrounding galaxies (and to select the 5th nearest neighbour),
\citet{Verley07a} take into account only galaxies of similar-size:
\begin{equation}\label{two}
 1/4 \times D_i < D_t < 4 \times D_i,
\end{equation}
where $D_i$ is the size (i.e. the optical diameter, which is the major axis at
an isophotal level of 25 mag/arcsec$^2$ in the B-band) of the candidate
isolated galaxy, and $D_t$ is the size of the galaxy $t$ from the
neighbour-sample.  The local number density of neighbour galaxies
provides a description of the environment in the vicinity of each
AMIGA galaxy, but does not take into account the mass (or size) of the 
perturbers explicitly. Therefore, as an alternative isolation
criterion, \citet{Verley07a} estimate the tidal strength affecting each
AMIGA galaxy, a parameter initially proposed by
\citet{Dahari84}. In particular, they compute $Q_{it}$, that is the ratio
between the tidal force and the binding force, which provides a measure for the
tidal strength a galaxy \textit{i} experiences by a neighbour galaxy
\textit{t}:
\begin{eqnarray}\label{three}
Q_{it} = \frac{F_{\mathrm{tidal}}}{F_{\mathrm{bind}}} \propto
\frac{M_t \times D_i}{S_{it}^3} \times \frac{D_i^2}{M_i},
\end{eqnarray}
In the above equation, $S_{it}$ is the projected separation between
the isolated galaxy $i$ and the neighbour galaxy $t$, and $M_i$ and
$M_t$ are the masses of the isolated and the neighbour galaxies,
respectively. Finally, $Q = \log (\sum_{t} Q_{it})$ is a dimensionless
estimation for the interaction strength. Also in this case, only
galaxies of similar size are considered according to
eq. \ref{two}. \citet{Verley07a} find that the majority of galaxies in
the AMIGA sample have values of$\log \Sigma_{5NN} < 2.5$ and/or $Q <
-2$. In the following analysis, we will use  these limits to select isolated
galaxy samples from our model galaxy catalogues.

%*******************************************************************************
%*******************************************************************************
\section{The theoretical framework}\label{theory}
%*******************************************************************************
%*******************************************************************************

%*******************************************************************************
\subsection{The galaxy formation model}\label{model}
%*******************************************************************************

In this study, we take advantage of the publicly available catalogues from the
galaxy formation model presented in \citet{DeLucia07}. This model was applied
to the dark matter merger trees extracted from the Millennium Simulation
(\citealp{Springel05}). The simulation assumes a WMAP1 cosmology with
$\Omega_{\Lambda} = 0.75$, $\Omega_{m} = 0.25$, $\Omega_{b} = 0.045$, $n=1$,
$\sigma_8 = 0.9$ and, $h=0.73$.  Note that more recent
  measurements of the CMB e.g. with the Planck satellite
  (\citealp{Planck13}) obtain a slightly smaller value for $\sigma_8 =
  0.83$. If only $\sigma_8$ is changed, the present-day Universe
  corresponding to a $\sigma_8$ lower than that used in our
  simulation can be well approximated by a snapshot corresponding to
  some earler redshift (see e.g. \citealp{Wang08}). The quantitative
  results presented in the following (e.g. the fraction of isolated
  galaxies) would change. However, we do not expect the qualitative
  trends discussed below to be altered significantly. The galaxy
formation model includes prescriptions for gas cooling, re-ionization,
star formation, supernova feedback, metal evolution, black hole
growth, and AGN feedback. For more details on these prescriptions, we
refer the reader to \citet{DeLucia07} and \citet{Croton06}. The model
has been shown to reproduce qualitatively a large variety of data,
both at high redshift and in the local Universe. However, it is not
without problems: the model predicts an excess of low-mass galaxies,
and the predicted fraction of red galaxies is too high with respect to
observational measurements. These drawbacks are not specific of this
galaxy formation model, rather represent one of the major challenges
even for recently published semi-analytic models
(e.g. \citealp{Bower06, Monaco05, Somerville08, Hirschmann12}), as
well as for hydrodynamic simulations (\citealp{Dave11,
  Weinmann12}). There have been some recent attempts to solve these
problems by assuming stronger SN feedback (\citealp{Guo11}) and/or by
changing the SF prescription (\citealp{Wang12}), but none of the
proposed solution is completely satisfactory and successful. 

We note, however, that the drawbacks just discussed do not affect
significantly the results presented in the following. In this respect,
\citet{Croton08} have investigated the red/blue void luminosity
functions for the same model and found them to be in reasonably good
agreement with the 2dFGRS results. We have explicitly verified that
our conclusions do not vary if we use the public catalogues from the
model discussed in \citet{Guo11}, applied to the Millennium II
simulation. Thus, we are confident that our results, based on a
comparison between isolated and non-isolated samples, are robust, even
if  our adopted model over-predicts in absolute terms the number of
low mass galaxies.

\begin{figure}
\begin{center}
 \epsfig{file=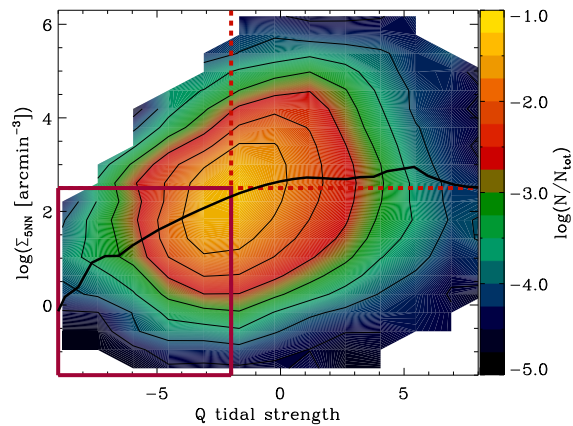,width=0.45\textwidth}
  \caption{2D histogram of the projected density computed using the fifth
    nearest neighbour $\Sigma_{5NN}$, versus the tidal strength parameter
    $Q$. The thick black line shows the median relation. Red lines indicate the
    cuts used for selecting isolated {\bf model} galaxies following
    the observational criteria by \citet{Verley07a}. The lower left
    purple square illustrates the selection what we call the Iso\_NN+Q
    sample, while the upper right region defines our high density
    sample.} {\label{Dens1_dens2}} 
\end{center}
\end{figure}

%*******************************************************************************
\subsection{Selecting isolated galaxies}\label{selection}
%*******************************************************************************

For selecting isolated galaxies from model catalogues, we will
consider two different kind of criteria:
\begin{enumerate}
\item{three 2D criteria, that mimic the observational selection. In
  particular, we will use the same criteria as in the study of
  \citet{Verley07a} (see description in section \ref{AMIGA}).}
\item{one 3D-criterion, where the three-dimensional information of model galaxies
  in the \textit{real space} will be used. This is accessible in simulations,
  but generally not in observational studies.}
\end{enumerate}
In Table~\ref{tab_selcrit}, we summarise the different isolated galaxy samples
that we will analyse in this work.

\begin{table*}
\caption{List of the different isolation criteria and the
  corresponding isolated galaxy samples we will analyse in this paper.}
\begin{tabular}{p{2.2cm}p{8.3cm}p{2.5cm}p{2.cm}}
\hline\\\bf{Name} & \bf{Criterion} & \bf{Type} & \bf{Colour code}\\ 
\hline \hline 
Iso\_NN & Local number density $\log \Sigma_{5NN} < 2.5$ & 
2D-observational & green\\
Iso\_Q & Tidal strength $Q<-2$ & 2D-observational & purple \\ 
Iso\_NN+Q &  Local number density $\log \Sigma_{5NN} < 2.5$ \& Tidal
strength $Q<-2$ & 2D-observational & orange\\ 
Iso\_3D$_{\mathrm{RS}}$ & No galaxy within a 1 Mpc sphere & 
3D-real space & cyan\\
\hline
\end{tabular}
\label{tab_selcrit}
\end{table*}

%*******************************************************************************
\subsubsection{2D selection}\label{obssel}
%*******************************************************************************

In order to select isolated galaxies in models following
observational definitions of isolation, we mimic a real observation by
putting a virtual observer in the middle of our simulation box at
$z=0$, and by converting the x, y, and z coordinates into right
ascension, declination and observed redshift (recessional velocity),
respectively. The latter includes the contribution from peculiar
velocities. Note that we assume $z \approx 0$ throughout the box for
our calculations. Following the AMIGA selection, we exclude nearby
galaxies with recessional velocities below $1,500\ \mathrm{km/s}$.  In
addition, like \citet{Verley07b}, we distinguish between a sample of
candidate isolated galaxies (this is our `parent sample', and includes
all galaxies down to $m_{b,j} = 15.7$), and a sample of neighbour
galaxies (the `neighbour-sample', with an apparent magnitude cut of
$m_{b,j} < 17.5$). Apparent magnitudes are calculated using the
absolute magnitudes available from the catalogues, and the luminosity
distances corresponding to the position of the galaxies within the box
with respect to the position of the virtual observer. 

We select isolated model galaxies by computing $\Sigma_{5NN}$
and $Q$ as outlined in section \ref{AMIGA}. As the model used does not
include prescriptions for size growth, we use the observed mass-size
relation published in \citet{Auger10} to assign a physical size to
each model galaxy. This is then converted into an apparent size using
the assumed geometry. 

Fig.~\ref{Dens1_dens2} shows a distribution of our model galaxies
(normalised to the total number of galaxies) as a function of the
tidal strength parameter $Q$, and of the projected local density
$\Sigma_{5NN}$. The entire parent sample contains roughly 65,000
galaxies, and about one third of them are satellites. We find only a
weak correlation between the tidal strength and the local density, 
and the majority of galaxies have a tidal strength $Q \sim -2$
and a local number density $\log \Sigma_{5NN} \sim 2$. The red lines in
Fig. \ref{Dens1_dens2} indicate the two isolation criteria we use in
the following analysis: (i) $\log \Sigma_{5NN} < 2.5$ and (ii) $Q<-2$
(consistent with the result of \citealp{Verley07a}). In the following,
we will consider three samples of 2D-isolated galaxies satisfying
either the first (i.e. `Iso\_NN' sample) or the second criterion
(i.e. `Iso\_Q' sample), or both criteria (i.e. `Iso\_NN+Q' sample, see
Table \ref{tab_selcrit}). 

%*******************************************************************************
\subsubsection{3D-real space selection}\label{3Dsel}
%*******************************************************************************

In contrast to observations, theoretical simulations provide the full
three-dimensional information in the real space for all galaxies. This allows a
`3D-real space' definition of isolation: we assume that a galaxy from the
parent sample is isolated only when no other galaxy (from the neighbour-sample)
is found within a sphere of co-moving radius of $1\ \mathrm{Mpc}$\footnote{Note
  that 1~Mpc is roughly the distance to the first nearest neighbour when
  applying both 2D-observational selection criteria. In addition, this
  selection results in a similar number of galaxies as in the Iso\_NN+Q
  sample.}. To provide a fair comparison between our different isolated galaxies
samples, we use the same parent sample for the 3D-real space as for the
2D-observational selection, i.e. we consider only galaxies with $m_{b,j} <
15.7$. As for the neighbour-sample, we consider only galaxies
with masses larger than one order of magnitude below the mass of the
isolated galaxy, i.e. $\log(M_{\mathrm{iso}}/M_{\odot})-1 <
\log(M_{\mathrm{neighbour}}/M_{\odot}) $. This way, we put a homogeneous
upper limit to the tidal strength exerted by the neighbours on each
isolated galaxy, irrespectively of its mass.

%*******************************************************************************
%*******************************************************************************
\section{Present-day galaxy properties}\label{galprop}
%*******************************************************************************
%*******************************************************************************

In this section, we focus on basic properties of present-day isolated model
galaxies selected by using different criteria. For each sample of isolated
galaxies, we construct a corresponding \textit{mass-matched random sample}
i.e. a random galaxy sample with the same magnitude cut, same total number of
galaxies, and the same stellar mass distribution of the corresponding isolated
sample. This allows us to take out the strong dependency of galaxy properties
on stellar mass, and perform an unbiased comparison between isolated
galaxies properties and `average' galaxy properties. 

%*******************************************************************************
\subsection{2D isolation criteria}\label{galprop_obs}
%*******************************************************************************

%*******************************************************************************
\subsubsection{Stellar masses}\label{Stellar_obs}
%*******************************************************************************
\begin{figure}
\begin{center}
 \epsfig{file=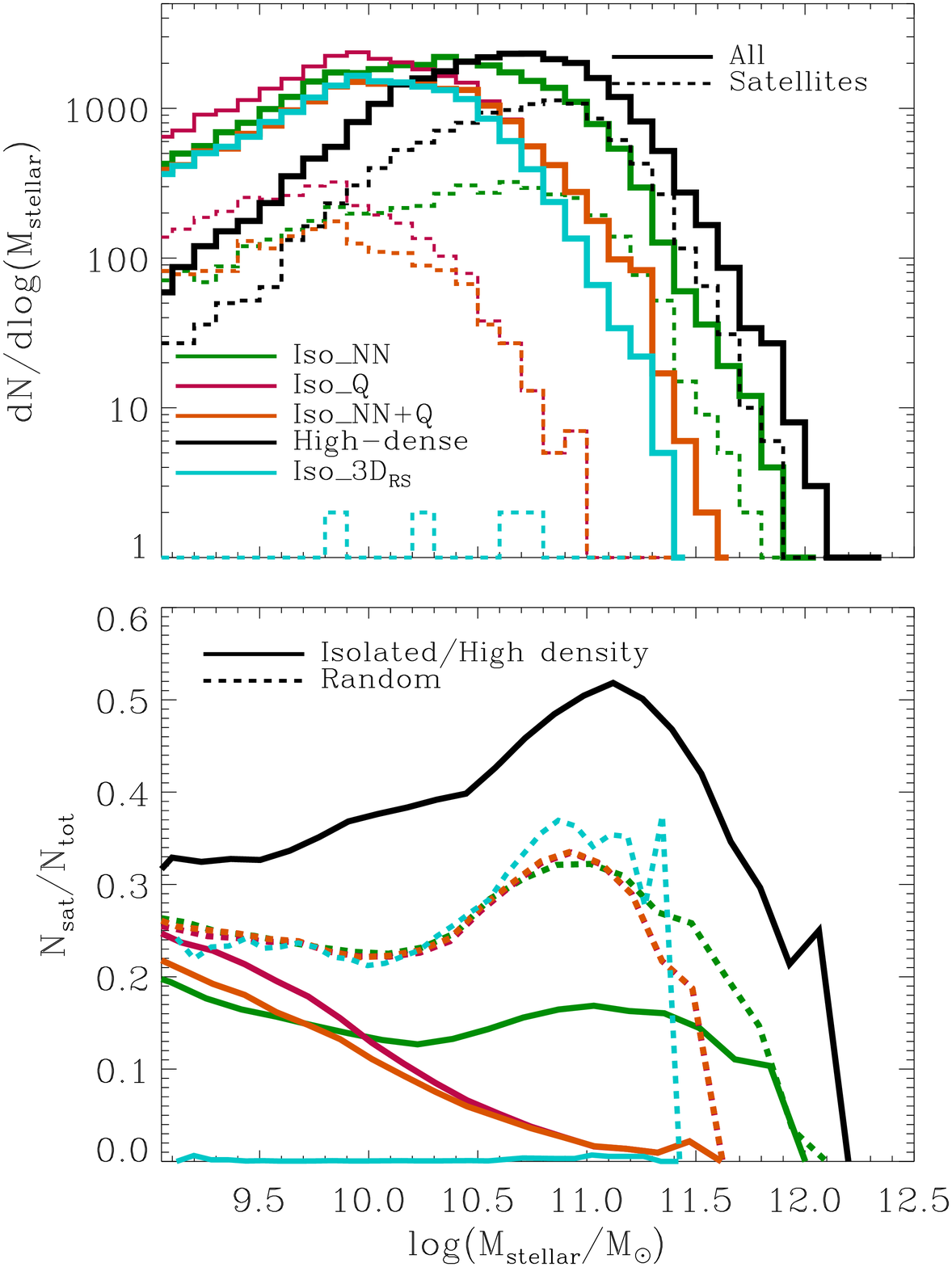,width=0.4\textwidth}
 \caption{Upper panel: present-day stellar mass distributions for 2D
   observationally and 3D-real space selected isolated {\bf model}
   galaxies (Iso\_NN: 
   green, Iso\_Q: purple, Iso\_NN+Q: orange, Iso\_3D$_{\mathrm{RS}}$:
   cyan). For comparison, the black line illustrates the corresponding
   distribution for `high-density' galaxies.  Dashed lines indicate the
   corresponding distributions for the satellite galaxies in each
   sample. Bottom panel: Fraction of satellite galaxies as a function of
   stellar mass for the different isolated samples, and the
   high-density sample considered in the upper panel.}
 {\label{Stellarmass}} 
\end{center}
\end{figure}

The top panel of Fig. \ref{Stellarmass} shows the stellar mass
distributions for the three 2D selected isolated model galaxy
samples, as indicated in the legend. For comparison, the black line
shows the stellar mass distribution of high-density galaxies residing
in the top right quadrant of Fig. \ref{Dens1_dens2} (i.e. those
corresponding to $\log \Sigma_{5NN} > 2.5$ and $Q > -2$). In
qualitative agreement with observational findings, isolated 
  model galaxies tend to have lower stellar masses than galaxies
residing in dense regions.  Comparing the different isolated samples
among each other shows that the stellar mass distributions of isolated
galaxies from the Iso\_Q and the Iso\_NN+Q samples are very close to
each other at the high mass end, and only slightly different at the
low mass end. The latter difference is likely due to galaxies that are
surrounded by a large number of lower/similar mass ones, failing the
Iso\_NN criterion but passing the Iso\_Q one.  Compared to the Iso\_NN
sample, galaxies from the Iso\_Q 
and from the Iso\_NN+Q samples tend to have significantly lower
stellar masses ($<10^{11.5} M_{\odot}$). In other words, when
selecting isolated galaxies using the Q-parameter, one loses a
relatively large number of massive galaxies that have a low numbers of
neighbours, but high Q values. These galaxies are preferentially living
as pairs, triplets or dense small systems: i.e. massive galaxies with
several massive neighbours close enough to `disturb' them
significantly. This shows the importance of adding also the Q-cut
(i.e. considering also the mass of the neighbouring galaxies) to avoid
`contamination' of isolated samples from these systems.

%*******************************************************************************
\subsubsection{Isolated satellite galaxies}\label{Isosat}
%*******************************************************************************

The bottom panel of Fig. \ref{Stellarmass} shows the fraction of
satellite model galaxies for the different isolated galaxy
samples. We find, independently of the isolation criterion used, a
clear prevalence of central galaxies for the isolated galaxy
population. Satellite galaxies constitute only a minor fraction of the
isolated galaxies (less than 15 per cent). A similar result was
  found for void galaxies by \citealp{Croton08} (see their Figure 2),
  although these galaxies are usually defined using large scale density
  estimators (e.g. 8~Mpc) in constrast to our selection of isolated
  galaxies. For comparison, the black line shows the satellites
fraction for the high-density galaxy sample (top right quadrant in
Fig. \ref{Dens1_dens2}), which is much larger than the corresponding
fraction for all isolated galaxy samples. However, when comparing the
2D isolated samples with the corresponding mass-matched random samples
(dashed lines of the same  colour), `typical' galaxies of the same
stellar mass have a significantly larger fraction of satellites ($\sim
25$ per cent) peaking at stellar masses about $10^{11} M_{\odot}$).
While for the Iso\_Q and the Iso\_NN+Q samples, 
the satellite fraction is decreasing with increasing stellar mass, for
the Iso\_NN sample the satellite fraction is roughly constant as a
function of stellar mass and equal to about 15 per cent. This confirms
that the high mass tail rejected by the Iso\_Q criterion is composed
by relatively massive galaxies located in small, isolated systems. In
these systems such galaxies would be the satellites.

\begin{figure}
\begin{center}
\epsfig{file=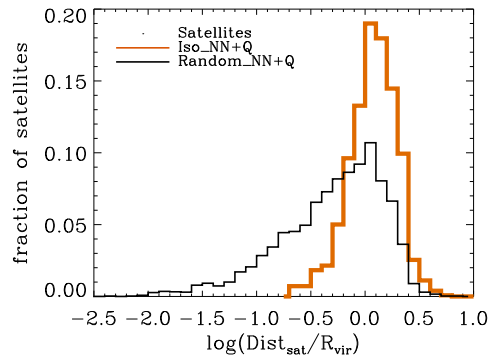,width=0.4\textwidth}
 \caption{Fractions of satellites versus the distances to their parent halo
   centre, normalised to the virial radius. The thick orange line corresponds
   to galaxies selected from the Iso\_NN+Q sample, while the thin black line
   illustrates the corresponding random sample.}  {\label{Raddist}}
\end{center}
\end{figure}

The presence of satellite galaxies in the isolated samples might appear
counter-intuitive, and certainly in contradiction with the term
`isolated'. Indeed, by definition, these galaxies have experienced
nurture-related physical processes at least for some fraction of their
lifetime. It is, therefore, interesting to understand why and under which
conditions, a satellite galaxy ends up being selected as a candidate isolated
galaxy. Fig. \ref{Raddist} shows the distribution of the radial distances of
satellites to their parent halo centre for the Iso\_NN+Q sample, and for the
corresponding mass-matched random satellite sample. The radial distances are
normalised to the virial radius of the parent halo. We find that isolated
satellites are most likely located at the outskirts or even outside of the
virial radii of the parent halos: $0.5 R_{\mathrm{vir}} <
\mathrm{Dist}_{\mathrm{sat}} < 3 R_{\mathrm{vir}}$.  In contrast, randomly
selected satellites tend to live closer to the parent halo centre and thus,
their distances span a broader range between $0.01 R_{\mathrm{vir}} <
\mathrm{Dist}_{\mathrm{sat}} < 3 R_{\mathrm{vir}}$. 

In summary, isolated samples selected using the criteria illustrated
in the section \ref{obssel} contain a fraction of satellite galaxies
ranging between 10 and 15 per cent, with the fraction being smaller
when the Q parameter is employed for the selection. The satellites
selected as isolated are generally those residing at the outskirts of
haloes, where the local galaxy density is lower (note that this is
largely a consequence of our definitions of isolation). Surprisingly,
we also find that roughly half (55 per cent) of the satellites
classified as isolated, have been satellites for more than 2
Gyr. Therefore, these galaxies have been exposed to environmental
physical processes, and their physical properties have not been
determined by internal physical processes only. We find that these
galaxies reside preferentially in massive host halos ($> 10^{13}
M_\odot$), and have been accreted with low average mass ratio
($M_{\mathrm{subhalo}}/M_{\mathrm{hosthalo}} \sim 0.02$), so that
their survival time as satellites are relatively long.   

The selection criteria used for our 2D sample are aimed at mimicking
the selections applied for the AMIGA sample. Fig. \ref{Dens1_dens2}
suggests that the Iso\_NN+Q sample is dominated by galaxies lying in
the central region of the $\Sigma_{5NN}$-$Q$-plane. To test how this
affects our results, we have considered an additional criterion that
adopts the following cuts: $Q<-4$ and $\log(\Sigma_{5NN})<1$. As
expected, the number of isolated model galaxies is reduced by using
this ``stricter'' criterion. We also find that it tends to select
galaxies of lower mass than the ``weaker'' criterion used above with a
lower fraction of satellites. 

%*******************************************************************************
\subsubsection{Morphology}\label{Morph_obs}
%*******************************************************************************

\begin{figure}
\begin{center}
 \epsfig{file=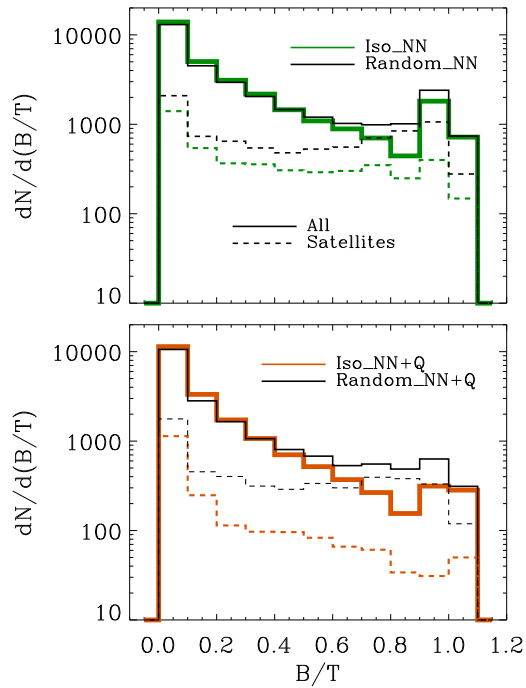,width=0.4\textwidth}
 \caption{Histograms of the bulge-to-total ratios for observationally selected
   isolated {\bf model} galaxies, and for the corresponding
   mass-matched random samples, as indicated in the legend. Dashed
   lines show the corresponding distributions 
   for the satellite galaxies. Different panels illustrate the different
   criteria (see legend).}  {\label{BT_obs}}
\end{center}
\end{figure}

In Fig.~\ref{BT_obs}, we show the distributions of the stellar mass
bulge-to-total ratios for two different 2D-isolated galaxy samples
(top panel: Iso\_NN; bottom panel: Iso\_NN+Q). We do not
discuss here the Iso\_Q sample separately, as results do not differ
significantly from those shown for the Iso\_NN+Q sample. The black
lines correspond to the respective mass-matched random sample, and
dashed lines illustrate the contribution from satellites. We find that
all isolated and random samples in our models are clearly
dominated by late-type galaxies, but with a varying fraction of
early-type (bulge-dominated) galaxies. Comparing the two isolated
samples, the Iso\_NN sample contains roughly $\sim 20$ per cent of
bulge-dominated (B/T ratio $ > 0.8$) galaxies, while the Iso\_NN+Q
sample includes only about 4 per cent of these galaxies. This can be
explained in part by the stellar mass distributions: compared to the
Iso\_NN sample, the Iso\_NN+Q galaxies are shifted towards lower
stellar masses, which have generally a smaller relative bulge mass. By
comparing isolated samples to the corresponding mass-matched random
ones, we find that the latter include a larger fraction of
bulge-dominated systems. The dominance of late-type galaxies among
isolated model galaxies is also in qualitative agreement with results
for void galaxies (see  e.g. \citealp{Croton08} based on 2dFGRS data
and \citealp{Kreckel12} based on SDSS data).  

It is interesting to study the sub-sample of isolated model
ellipticals (with B/T ratio $ > 0.8$) in more detail. We find that it
contains only few low-mass galaxies, i.e. isolated ellipticals tend to
reside among the most massive in each sample considered,  with a peak
in their stellar mass distributions between $10^{10.5}$ and $10^{11}
M_{\odot}$. In the galaxy formation model considered in our study,
bulges can grow through merger events and through secular evolution
processes like disk instabilities.  Distinguishing the contribution
from these two channels, we find that bulges of isolated early-type
galaxies (Iso\_NN+Q) have been built either by disk instabilities
only or by merger events only, but not through a combination of these 
processes. The vast majority of these bulges have been formed via
merger events with a mass ratio between 1:1 and 1:10. In addition, we
find that roughly two-thirds of the merger-formed bulges of isolated
ellipticals have been assembled via minor mergers (mass ratio between
1:4 and 1:10), just like for the respective mass-matched random sample.

The presence of bulge-dominated galaxies in the isolated samples, and the
finding that most of these bulges are formed through merger events, suggest
that at least a fraction of the present-day isolated galaxies (the majority of
those with a high bulge-to-total ratio) have not always been isolated during
their lifetime, and must have been interacting with neighbouring galaxies. This
result will be discussed in more detail in section \ref{isomerger}. We note
that our findings are in agreement with the study by \citet{Niemi10}. They used
galaxy catalogues from the same model adopted in this study, and found 
that almost all ($\sim$98 per cent) of the isolated ellipticals experience some
merger activity during their evolution. However, while about one third of
\textit{our} isolated early-type galaxies have experienced at least one major
merger event during their lifetime, \citet{Niemi10} find a larger fraction of
isolated ellipticals ($\sim$46 per cent) having experienced at least one major
merger. The discrepancy is likely caused by the different magnitude cuts and
isolation criteria used.

%*******************************************************************************
\subsubsection{Star formation}\label{star_form}
%*******************************************************************************

\begin{figure}
\begin{center}
 \epsfig{file=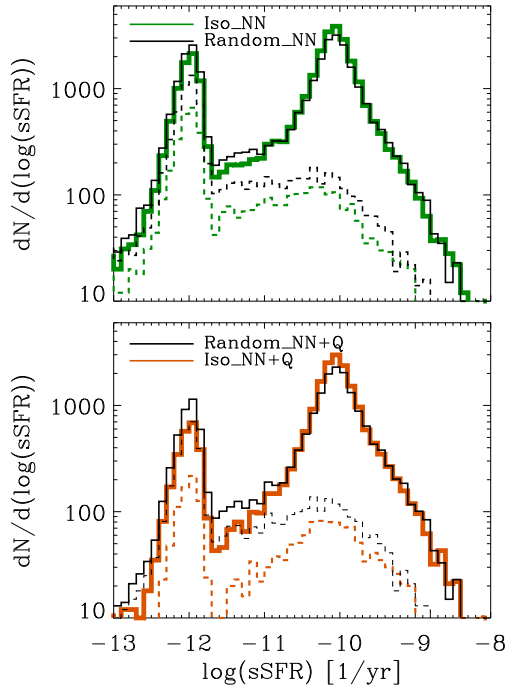,width=0.4\textwidth}
 \caption{Histograms of the specific star formation rates for
   2D-observationally selected isolated {\bf model} galaxies, and for
   the corresponding mass-matched random samples. Dashed lines show
   the corresponding distributions for satellite galaxies. Different
   panels illustrate the different selection criteria, as indicated in
   the legend.}{\label{sSFR_obs}} 
\end{center}
\end{figure}

Fig. \ref{sSFR_obs} shows the distribution of the Specific Star
Formation Rates (sSFR = SFR$/M_{\mathrm{stellar}}$) for two different
2D-isolated galaxy samples (top panel: Iso\_NN; bottom panel:
Iso\_NN+Q). The black lines correspond to the respective mass-matched
random sample, while dashed lines illustrate the contribution from
satellite galaxies. We find that all isolated and mass-matched random
samples are dominated by star forming galaxies (with $\log
\mathrm{sSFR} > -11$), which is a direct consequence of the fact that
isolated galaxies (in all samples) are dominated by galaxies of low
stellar masses. In the Iso\_NN sample, roughly 70 per cent of the
galaxies are star-forming, and the distribution of isolated galaxies
is very similar to that of the corresponding mass-matched random
sample. In contrast, in the Iso\_NN+Q sample, a larger fraction of
galaxies are star-forming ($\sim 85\%$) than in the corresponding
mass-matched random sample ($\sim 68\%$). A predominance of
star-forming galaxies has also been found in theoretical and
observational studies focused on void galaxies (e.g. \citealp{Hogg03,
  Hoyle04, Croton05, Croton08, Hoyle12, Kreckel12} and references
therein). 

Independently of the galaxy sample, satellites provide a relatively
larger contribution to the quiescent fraction of isolated galaxies
than centrals. However, this is basically by definition. Indeed, the
galaxy formation model used in this study adopts an instantaneous
strangulation of the hot gas reservoir after accretion of galaxies
onto larger systems. This switches off cooling on satellite galaxies,
and quenches their star formation on very short time-scales. It is
known that this assumption produces an excess of faint and passive
satellites in galaxy formation models. As a consequence, different
studies (e.g. \citealp{Weinmann10, Wetzel12, DeLucia12} and references 
therein) suggest much longer quenching time-scales, with typical
values of $5-7$ Gyr. 

%*******************************************************************************
\subsubsection{Parent halo masses}\label{Halo_obs}
%*******************************************************************************

\begin{figure}
\begin{center}
 \epsfig{file=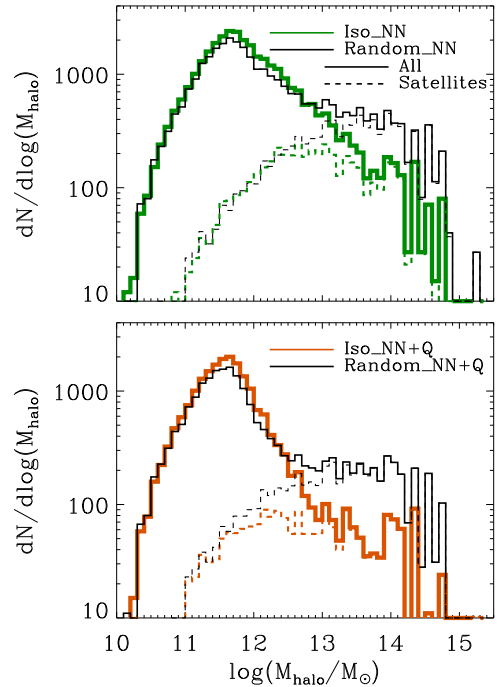,width=0.4\textwidth}
 \caption{Distributions of the parent halo masses for 2D-observationally
   selected isolated galaxies, and for the corresponding mass-matched random
   samples. Dashed lines show the corresponding distributions for the
   subsamples of satellite galaxies. Different panels illustrate the
   different selection criteria, as indicated in the
   legend.}{\label{Halomass_obs}}
\end{center}
\end{figure}

Fig. \ref{Halomass_obs} shows the distributions of parent halo masses
for two isolated model galaxy samples (top panel: Iso\_NN;
bottom panel: Iso\_NN+Q), and for the corresponding mass-matched
random samples. The majority of both random and isolated galaxies
reside in parent halos with masses below $10^{13} M_{\odot}$, and
these galaxies are most likely centrals. In contrast, the massive ends
of the distributions (above $10^{13} M_{\odot}$) are completely
dominated by satellite galaxies. For the Iso\_NN sample, we obtain a
slightly larger fraction of isolated galaxies living in halos with
masses above $10^{12} M_{\odot}$ than for the Iso\_NN+Q sample. This
is just a reflection of the larger amount of (massive) satellite
galaxies in the Iso\_NN sample (see bottom panel in
Fig. \ref{Stellarmass}), which are know to preferentially reside in
massive halos. 

Comparing the isolated to their mass-matched random samples, we find
that the latter exhibit a significantly larger amount of galaxies
residing in halos above $10^{13} M_{\odot}$. Again, this is a direct
consequence of the larger fraction of satellites in the mass-matched
random samples than in the isolated ones. At the low mass end (below
halo masses of $\sim 10^{13} M_{\odot}$) the halo mass distributions
of the isolated and random samples are very close to each other,
independently of the chosen isolation criterion. Vice versa, the
difference at the high mass end between the mass-matched random and
isolated samples is stronger for the Iso\_NN+Q  sample. Our results
imply that the probability for an isolated galaxy to reside in a
massive halo is significantly reduced compared to a typical,
non-isolated galaxy of the same mass (because the galaxy population of
massive haloes is dominated by satellite galaxies, and an isolated
galaxy has a lower probability of being a satellite).
 
\begin{figure}
\begin{center}
 \epsfig{file=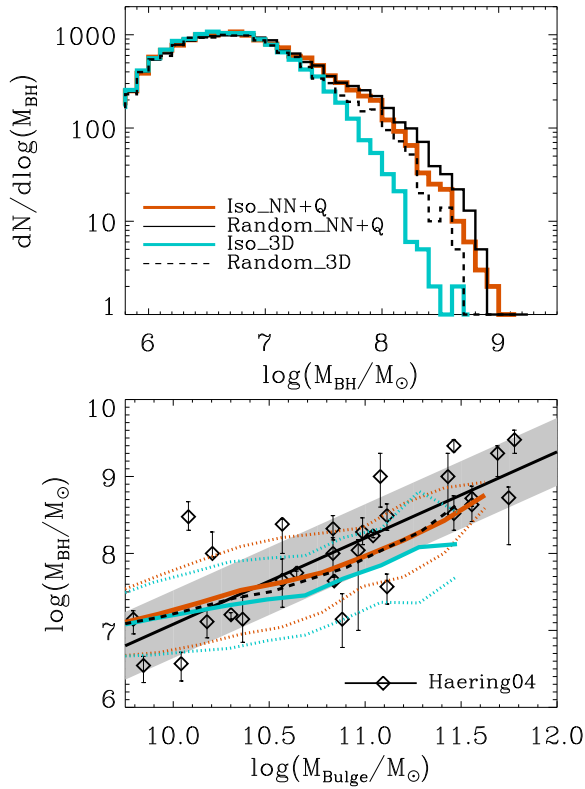,width=0.45\textwidth}
 \caption{Top panel: Distributions of the black hole masses for the Iso\_NN+Q
   and the Iso\_3D$_{\mathrm{RS}}$ isolated galaxies, and for the corresponding
   random samples. Bottom panel: Black hole-bulge mass relation for the
   Iso\_NN+Q, the Iso\_3D$_{\mathrm{RS}}$, and the Random\_3D$_{\mathrm{RS}}$
   samples (the relation corresponding to the Random\_NN+Q sample is identical
   to that of the Iso\_NN+Q one). Dotted lines indicate the 1-$\sigma$
   scatter. Model predictions are compared with the observed relation from
   \citet{Haering04} (black symbols). The black solid line with the grey shaded
   area shows a fit to the observational data.}  {\label{blackholes}}
\end{center}
\end{figure}

%*******************************************************************************
\subsubsection{Black holes in isolated galaxies}\label{BH_obs}
%*******************************************************************************

In our galaxy formation model, the formation of bulges through mergers proceeds
in parallel with the formation of black holes. As seen in section
\ref{Morph_obs}, a non negligible fraction of isolated galaxies do also host
a bulge, though this is generally not the dominant component. Hence, it is
interesting to analyse the relation between black hole and bulge masses of
isolated galaxies in order to investigate whether it deviates systematically
from that measured for the global bulge-dominated galaxy population. For the
Iso\_NN+Q sample and the corresponding mass-matched random one, we find that
roughly 75 per cent of the galaxies host black holes with masses above $10^6
M_\odot$, but they do not contain any black holes with masses above $10^9
M_\odot$ (see upper panel of Fig. \ref{blackholes}, orange and black solid
lines). The black holes of the Iso\_NN+Q sample are slightly less
massive than the ones of the corresponding mass-matched random sample. Thus,
the fact that isolated galaxies do not host the most massive black holes in the
Universe is not just due to the fact that their stellar mass distribution is
skewed towards low values. 

The lower panel of Fig. \ref{blackholes} shows the black hole-bulge mass
relation for isolated galaxies in the Iso\_NN+Q as an orange line. We do not
explicitly show the relation for the mass-matched random sample as it is nearly
identical to the one obtained for the isolated sample. When comparing this
relation to the observed one (\citealp{Haering04}) for the overall galaxy
population (illustrated by the symbols and the black line with the grey shaded
area), we find that black holes with mass $>10^{10.2}M_\odot$ in isolated
galaxies tend to be slightly under-massive at a given bulge mass. This is,
however, true also for the randomly selected sample: in other words, the model
we are using predicts a black hole-bulge mass relation which is slightly
shallower than in the observations (see e.g. \citealp{Marulli08}).

In summary, the isolation criteria tends to exclude the most massive black
holes, but does not alter significantly the predicted black hole-bulge mass
relation with respect to that obtained from a random mass matched sample.

%*******************************************************************************
\subsection{3D-real space isolation criterion }\label{galprop_3D}
%*******************************************************************************

\begin{figure}
\begin{center}
 \epsfig{file=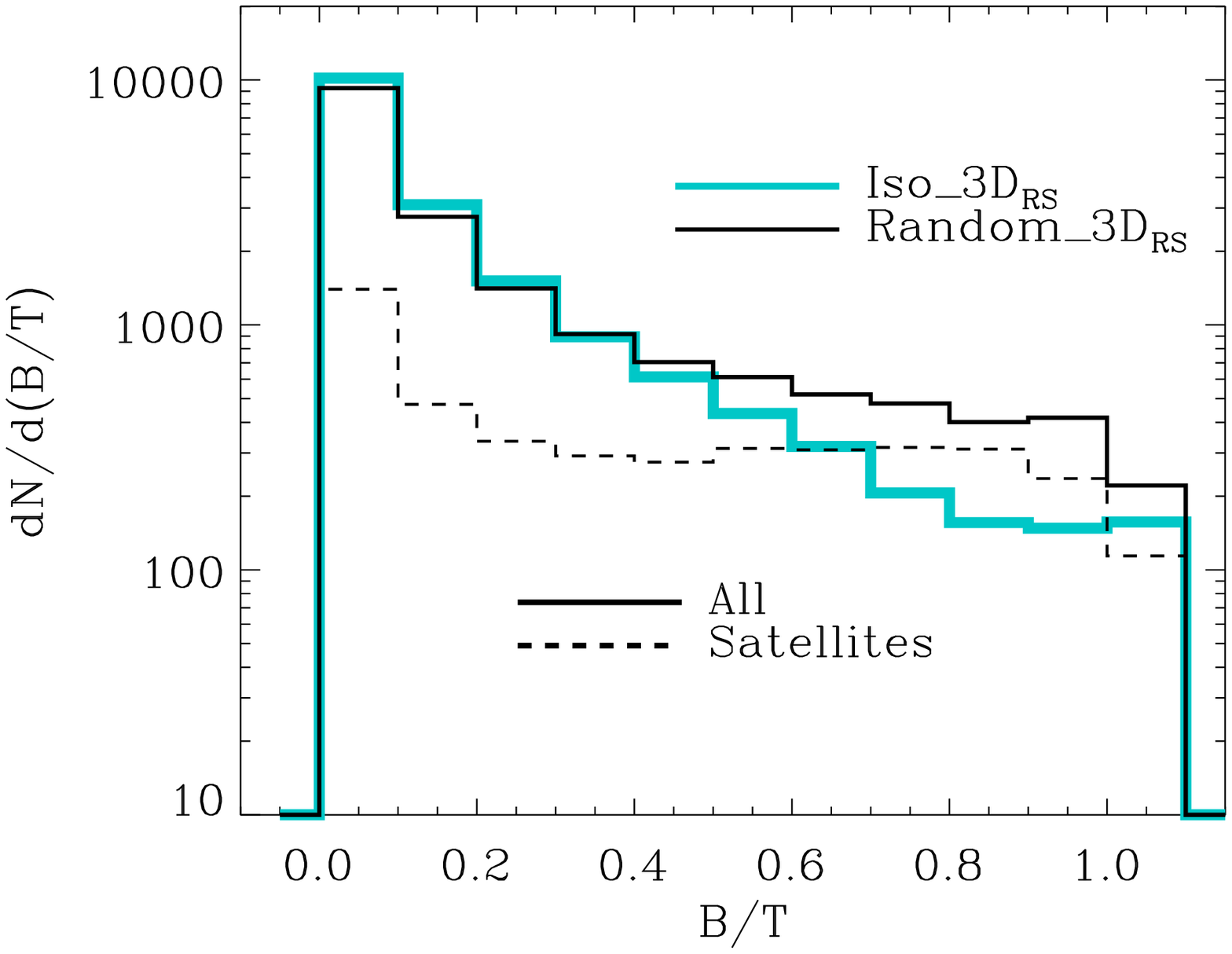,width=0.4\textwidth}
 \epsfig{file=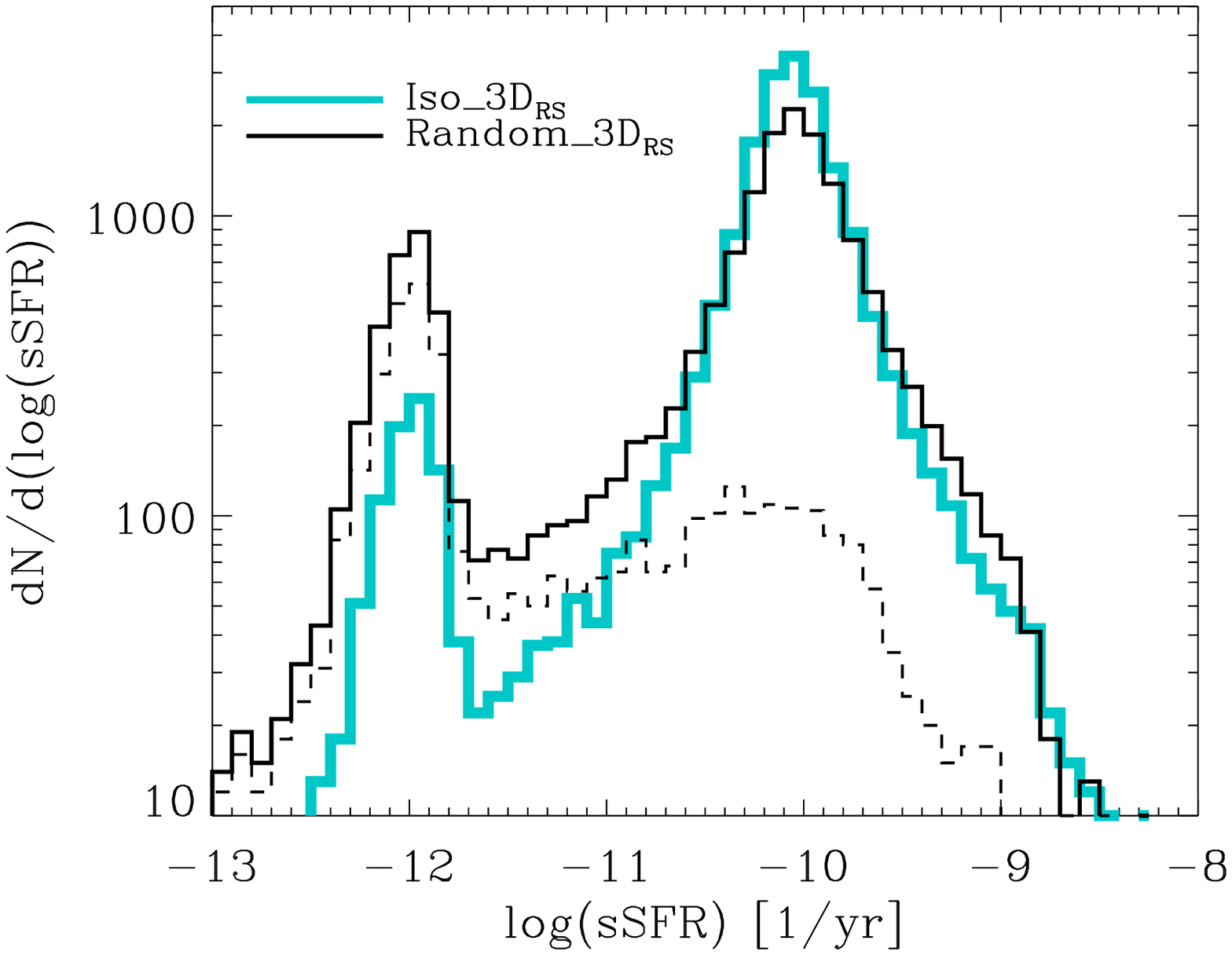,width=0.4\textwidth}
 \epsfig{file=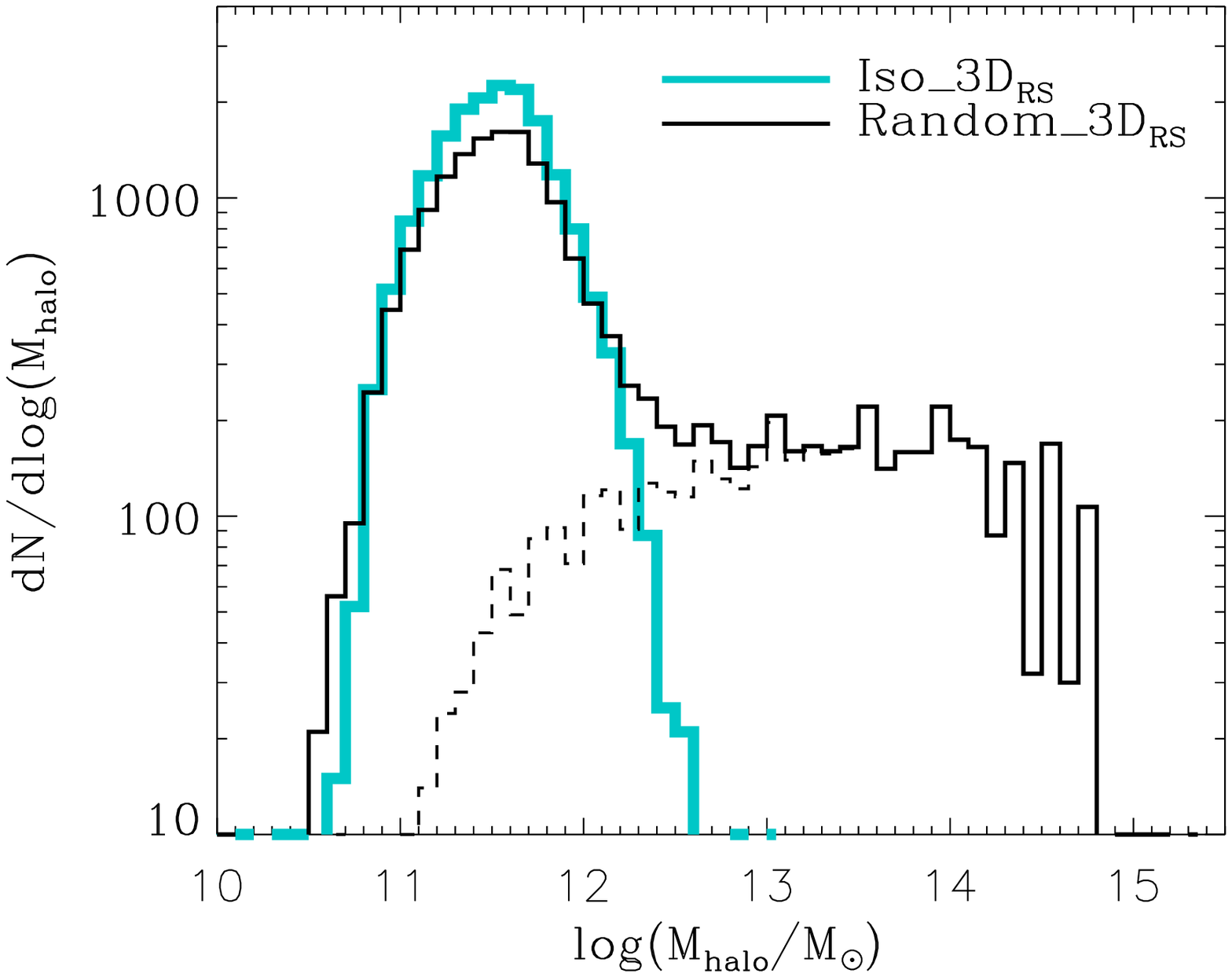,width=0.4\textwidth}
 \caption{Histograms of the bulge-to-total ratio (top panel), of the specific
   SFR (middle panel) and of the parent halo mass (bottom panel), using the
   3D-real space selected isolated galaxy sample (cyan line) and its
   mass-matched random sample (black line). Dashed lines illustrate satellite
   galaxies.} {\label{Propz0_ideal}}
\end{center}
\end{figure}

We now turn to the full 3-dimensional real space information from our
theoretical models, and focus on the 3D isolation criterion described
in section \ref{3Dsel}.  The cyan lines in Fig. \ref{Stellarmass} show
the stellar mass distribution (top panel) and the fraction of
satellite galaxies (bottom panel) for the isolated sample selected
using the full 3D information. In this case, isolated galaxies have
stellar masses lower than $3\times 10^{11} M_{\odot}$, and their mass
distribution shape is only slightly different than that of
Iso\_NN+Q. In addition, the Iso\_3D$_{\mathrm{RS}}$ sample contains 
only a few satellite galaxies: the 3D isolation criterion used is
quite efficient in removing the residual satellite galaxies entering
the Iso\_NN+Q sample, which - as we discussed in section \ref{Isosat}
- inhabit mostly the outskirts of their parent haloes. We note that
given the scale and mass limits adopted, this happens almost by
construction: excluding galaxies with neighbours with stellar mass down
to one order of magnitude below the candidate isolated galaxy, and
within 1 Mpc from it, excludes most of the satellite galaxies.

Fig. \ref{Propz0_ideal} shows the distribution of the bulge-to-total ratios
(top panel) for the Iso\_3D$_{\mathrm{RS}}$, and the corresponding mass-matched
random sample. Dashed lines illustrate the contribution from satellite
galaxies. Regarding the morphology, we find that the Iso\_3D$_{\mathrm{RS}}$
galaxies tend to be clearly less bulge-dominated than the corresponding
mass-matched randomly selected galaxies. This is mainly due to the negligible
amount of isolated satellite galaxies compared to the random sample, which tend
to be bulge-dominated (see the black dashed line). The same qualitative trend,
though less pronounced, was found for the Iso\_NN+Q sample (see
Fig. \ref{BT_obs}). As the Iso\_NN+Q sample, the
Iso\_3D$_{\mathrm{RS}}$ sample contains a small (about 2.7 per cent)
fraction of isolated early-types ($B/T>0.8$), which tend to be more
massive than a ``typical'' isolated galaxy. For 3D selected isolated
early-types, we find that about 86 per cent of the bulges have been
formed via major and minor mergers only, a slightly smaller fraction
than for the Iso\_NN+Q sample. Again, for these merger-formed isolated
early-types, minor mergers contribute two thirds of the mass of the
bulge component, on average. This again means that a small fraction of 
Iso\_3D$_{\mathrm{RS}}$ galaxies (all the isolated early-types) have
not been isolated for their entire life, but have been subject to
interactions with other galaxies. 

The middle panel of Fig. \ref{Propz0_ideal} shows the distribution of specific
SFRs for the Iso\_3D$_{\mathrm{RS}}$ and the corresponding
mass-matched random sample. We find that 3D$_{\mathrm{RS}}$ isolated
galaxies include a high percentage of star-forming galaxies ($\sim 94$
per cent), which is slightly higher than the one of the Iso\_NN+Q
sample, and significantly larger than the one of the corresponding
mass-matched random galaxy sample ($\sim 77$ per cent). A large part
of the difference between the Iso\_3D$_{\mathrm{RS}}$ sample and the
mass-matched random sample is due to the fact that the isolated sample
includes only an extremely low fraction of satellite galaxies, which
tend to be passive in our galaxy formation model. We find that roughly
all of the passive isolated galaxies in the Iso\_3D$_{\mathrm{RS}}$
sample are central galaxies with relatively high stellar masses ($\sim
10^{10.5-11.5} M_{\odot}$), likely quenched by AGN feedback.

Turning now to the parent halo mass distribution (bottom panel of
Fig. \ref{Propz0_ideal}), we find that 3D$_{\mathrm{RS}}$ isolated
galaxies are all living in haloes of mass below $10^{13}
M_{\odot}$\footnote{Note that this is in contrast to the results of
  \citet{Croton08}, but it may be due to the fact that they consider a
  criterion for void galaxies smoothed on much larger scales}. In
contrast, ``typical'' galaxies at various densities (the mass-matched
random sample) have a non-negligible probability to live in parent
halos with masses above $10^{13} M_{\odot}$, where the main
contribution comes from satellite galaxies missing in the
Iso\_3D$_{\mathrm{RS}}$ sample. When comparing the 2D selected
isolated galaxies (Iso\_NN+Q) with the Iso\_3D$_{\mathrm{RS}}$ ones,
we find that the amount of galaxies living in massive halos is
significantly larger in the Iso\_NN+Q sample. Again, as explained
earlier, this is due to the satellites living at the outskirts of
massive halos in the Iso\_NN+Q sample. 

Regarding black holes in 3D$_{\mathrm{RS}}$ isolated galaxies and in the
corresponding mass-matched random sample, Fig. \ref{blackholes} shows
the black hole mass distributions (cyan and black dashed lines in the upper
panel). 3D$_{\mathrm{RS}}$ isolated black holes have masses that do not exceed
$10^{8.5} M_\odot$, and are typically significantly less massive than the black
holes in the Iso\_NN+Q galaxy sample and in the Random\_3D
sample. Furthermore, considering the black hole-bulge mass relation
for the Iso\_3D$_{\mathrm{RS}}$ and Random\_3D$_{\mathrm{RS}}$ galaxy
samples (see cyan and black dashed lines in the bottom panel of
Fig. \ref{blackholes}), we find that both samples produce slightly
under-massive black holes at a given bulge mass compared to the
observed relation. However, the Iso\_3D$_{\mathrm{RS}}$ black holes
are even more under-massive than the ones of typical galaxies with the
same stellar mass, and also than Iso\_NN+Q black holes (orange
line). So in the case of the Iso\_3D$_{\mathrm{RS}}$, the isolation
criterion used affects both the number density of the most massive
black holes, and the median black hole-bulge mass relation. 

\begin{figure}
\begin{center}
 \epsfig{file=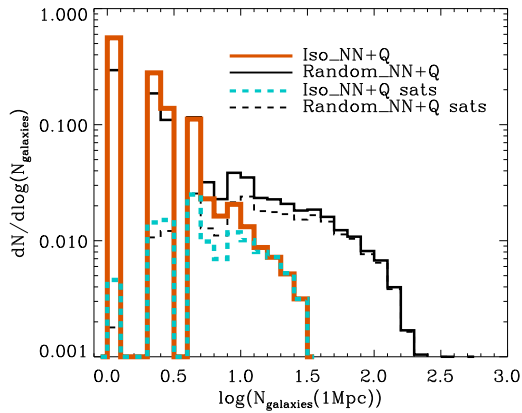,width=0.4\textwidth}
 \caption{Histogram showing the number of galaxies found within a
   1~Mpc sphere (including the central one) for the Iso\_NN+Q
   (orange), and for the corresponding mass-matched random 
   galaxy sample (black). Satellites are indicated by cyan and black dashed
   lines, respectively.}  {\label{Count1_obs}}
\end{center}
\end{figure}

%*******************************************************************************
\subsection{3D-isolation degree of 2D selected isolated galaxies }\label{iso_deg}
%*******************************************************************************

The differences between 2D and 3D$_{\mathrm{RS}}$ selected isolated
model galaxies are largely due to projection effects affecting the 2D
sample but also to the fact that they are defined in different
ways. Although our 3D criterion is also somewhat arbitrary  (one can
use a different radius and/or different mass limits), it is
interesting to quantify which fraction of the 2D selected isolated
galaxies (those in the Iso\_NN+Q sample) would also be isolated using
the 3D-real space information. Fig. \ref{Count1_obs} shows the
distribution of the number of galaxies within a 1~Mpc sphere for the
Iso\_NN+Q galaxies, and for the corresponding mass-matched random
sample. As usual, satellites are indicated by dashed lines. We find
that about 56 per cent of 2D isolated galaxies, and about 28 per cent
of the corresponding random galaxies are isolated when considering our
3D criterion (no neighbour galaxy within the 1~Mpc sphere with a mass
larger than one order of magnitude below the mass of the isolated
galaxy). However, there is a large percentage (about 82 per cent) of
the 2D isolated galaxies, which have less than 3 neighbours within the
1~Mpc sphere. The main contribution to 2D isolated galaxies (as well
as to the random galaxies) with more than 3 galaxies within the 1~Mpc
sphere is due to the satellite galaxies. Therefore, in a
2D-observationally selected galaxy sample, those that are
`misclassified' as isolated are largely the \textit{satellite}
galaxies, at least when considering our 3D$_{\mathrm{RS}}$ criterion.   

The `incompleteness' of the 2D\_NN+Q galaxy sample, i.e. the fraction of 3D
isolated galaxies that are not considered to be isolated due to projection
effects, can be quantified by calculating the fraction of
Iso\_3D$_{\mathrm{RS}}$ galaxies which do not satisfy our 2D-observational
criteria. We find that roughly 35 per cent of Iso\_3D$_{\mathrm{RS}}$ do not
satisfy the 2D-observational isolation criteria. The properties of the
galaxies that are selected as isolated using our 3D criterion are, however,
not significantly different from those that are selected as isolates using
the 2D criterion. Therefore, projection effects reduce the completeness of
isolated galaxy samples, but do not affect significantly the
distributions of their  physical properties.

\begin{figure}
\begin{center}
 \epsfig{file=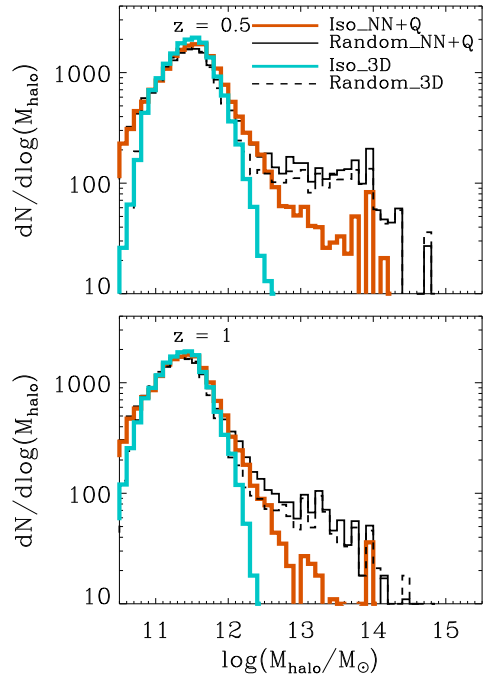, width=0.4\textwidth}
 \caption{Distributions of parent halo masses for progenitors of the Iso\_NN+Q
   (orange) and the Iso\_3D$_{\mathrm{RS}}$ (cyan) samples, and of their
   mass-matched random samples (black). Different panels refer to different
   redshifts (top: z=0.5, bottom: z=1).}  {\label{Halozevol_obs}}
\end{center}
\end{figure}
\begin{figure}
\begin{center}
 \epsfig{file=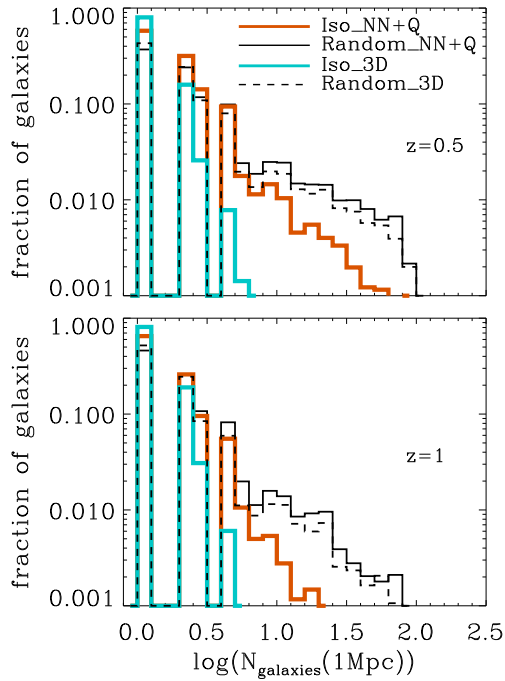, width=0.4\textwidth}
 \caption{Distributions showing the number of {\bf model} galaxies
   within a 1~Mpc sphere (including the central one) by using the full
   3D-real space information for progenitors of the Iso\_NN+Q 
   (orange) and of the Iso\_3D$_{\mathrm{RS}}$ (cyan) samples, and of their
   mass-matched random samples (black). Different panels refer to
   different redshifts (top: z=0.5, bottom: z=1).}{\label{Isoevol_obs}}
\end{center}
\end{figure}

%*******************************************************************************
%*******************************************************************************
\section{Environmental history of present-day isolated galaxies}\label{EnvHist}
%*******************************************************************************
%*******************************************************************************

As we have explained in section \ref{intro}, the interest in isolated
galaxies is driven by the fact that their physical properties are
believed to be mainly determined by internal physical processes
(i.e. by `nature'). Isolated galaxies can thus be used as a reference
sample for galaxies that are located in denser environments in order
to disentangle the relative importance of nature and nurture in galaxy
evolution. In this respect, it is important to understand to which
extent isolated galaxies have been isolated during their entire
life-time. In this section, we will study the `environmental history
of isolated model galaxies'. In particular, we will analyse their parent
halo mass distribution, and the number of neighbour galaxies within a 
sphere of 1~Mpc at different redshifts, and the redshift of the last
major/minor merger events that isolated galaxies have experienced. We 
will focus on the Iso\_NN+Q sample, the 3D-real space
Iso\_3D$_{\mathrm{RS}}$ sample, and the corresponding mass-matched
random galaxy sets. For this analysis we only consider the
\textit{main progenitors} of the present-day isolated/random galaxies,
i.e. the branch of the tree that is obtained by connecting the galaxy
to its most massive progenitor at each node of the tree.

%*******************************************************************************
\subsection{Halo masses}
%*******************************************************************************

Fig. \ref{Halozevol_obs} shows the parent-halo mass distribution for
the progenitors of present-day isolated galaxies (Iso\_NN+Q: orange;
Iso\_3D$_{\mathrm{RS}}$: cyan) and of the corresponding mass-matched
random samples (black lines), at $z=0.5$ and 1. Independently of the
galaxy sample considered, the high halo mass end is shifted towards
lower halo masses (i.e. the number of galaxies in massive halos
$M_{\mathrm{halo}} > 10^{13} M_{\odot}$ decreases) with increasing
redshift. This is a natural consequence of hierarchical structure
formation: small objects form first, and grow into larger
structures via smooth accretion and merger events. At all redshifts,
progenitors of the Iso\_3D$_{\mathrm{RS}}$ sample contain a smaller
number of galaxies residing in massive halos than the ones of the
Iso\_NN+Q sample. 

Comparing the Iso\_NN+Q or the Iso\_3D$_{\mathrm{RS}}$ to the
corresponding random galaxy samples, we find that the parent halo mass 
distribution is very close to that of isolated galaxies for low mass
haloes ($M_{\mathrm{halo}} < 10^{13} M_{\odot}$). This is not
surprising, given that these haloes dominate the halo mass function at
all redshifts. If any, we find that a larger fraction of the
progenitors of isolated galaxies are residing in low mass haloes with
respect to progenitors of galaxies from the random sample (this is
particularly evident for the Random\_NN+Q). At the high mass end, 
there is a significantly larger fraction of progenitors of the random
sample in massive halos with respect to progenitors of the isolated
galaxy sample. Again, this difference is mainly caused by the larger
amount of present-day satellite galaxies (which have been satellites
for some time in the past) in the random sample than in the isolated
sample.

\begin{figure}
\begin{center}
 \epsfig{file=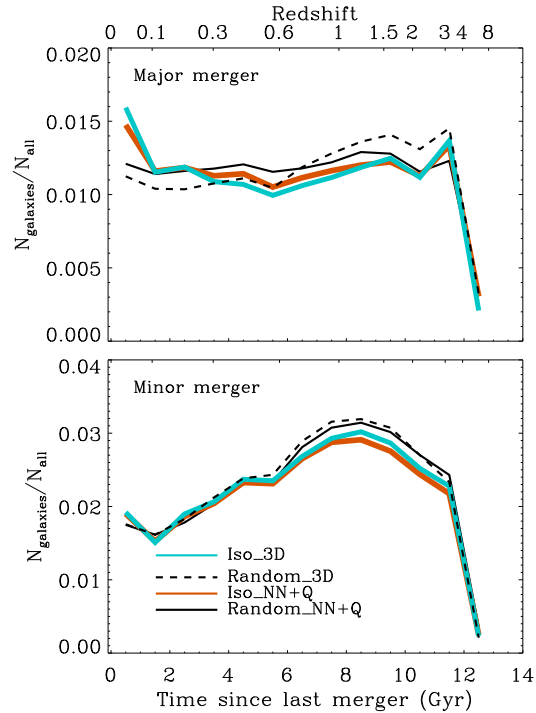, width=0.4\textwidth}
 \caption{Distributions of the lookback time since the last major (mass-ratio
   of 1:1-1:4, top panel) and minor merger (mass-ratio of 1:4-1:10, bottom
   panel) event. Coloured lines are for isolated samples (Iso\_NN+Q and
   Iso\_3D$_{\mathrm{RS}}$), and black lines for the corresponding mass matched
   random samples.}  {\label{LastMerger}}
\end{center}
\end{figure}
\begin{figure}
\begin{center}
 \epsfig{file=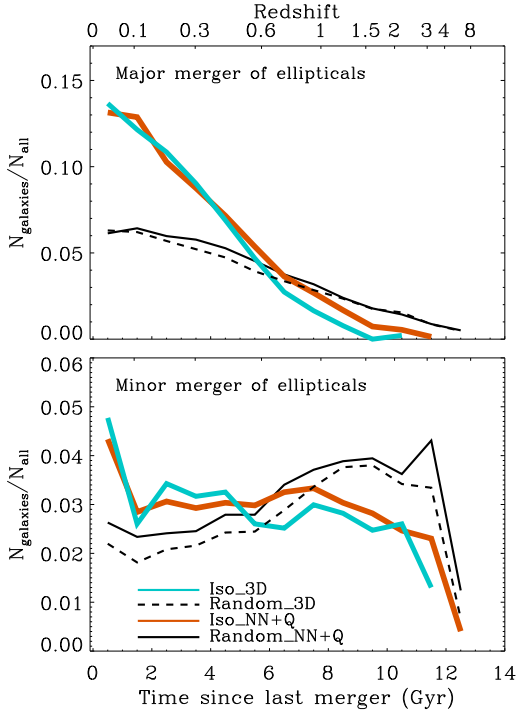, width=0.4\textwidth}
 \caption{Same as Fig. \ref{LastMerger}, but now for isolated and randomly
   selected ellipticals with B/T$>$0.8. Coloured lines are for isolated
   ellipticals (Iso\_NN+Q and Iso\_3D$_{\mathrm{RS}}$), while black lines for
   the corresponding mass matched random samples.} {\label{LastMerger_ell}}
\end{center}
\end{figure}

%*******************************************************************************
\subsection{Quantifying the degree of isolation since z=1 }
%*******************************************************************************

In this subsection, we ask the question whether present-day isolated
galaxies have been isolated for a significant fraction of their
life-time using the full 3D-real space information accessible in our
models. Fig.~\ref{Isoevol_obs} shows the distributions of the number
of neighbour-galaxies within a co-moving 1~Mpc sphere for the
progenitors of the 2D-isolated (Iso\_NN+Q), the 3D-isolated
(Iso\_3D$_{\mathrm{RS}}$) and the corresponding mass-matched random
samples at $z=0.5$ and 1. Note that we have used the same definition
of mass limits for neighbours at higher redshifts as at $z=0$ and the
3D selection. As the stellar mass limit for the neighbour sample
depends on the stellar mass of the parent galaxy, and the stellar mass
of the progenitors decreases with increasing redshift, the stellar
mass limit of the neighour sample will also vary (decrease) at higher
redshift. Our neighbour samples include a small fraction of galaxies
that are formally below the resolution limit of the simulation, but
this does not affect the results discussed here.   

We find that the amount of galaxies with a large number of neighbour 
galaxies within a 1~Mpc sphere slightly decreases with increasing
redshift, independently of the galaxy sample considered. For the
Iso\_NN+Q sample, about 60 per cent of present-day isolated galaxies 
had no other neighbour galaxy within a 1~Mpc sphere at z=0.5. The
fraction slightly increases to about 65 per cent at z=1. The same
fractions for the Iso\_3D$_{\mathrm{RS}}$ sample, are larger than for
the Iso\_NN+Q sample: at any redshift (since z=1), roughly 80 per cent
of present-day isolated galaxies have been isolated according to our
3D criterion. Therefore, Iso\_3D$_{\mathrm{RS}}$ galaxies have been on 
average `more isolated' during their life than Iso\_NN+Q
galaxies. This is not surprising as already at z=0 only 56 per cent of
Iso\_NN+Q satisfy the 3D-real space isolation criterion (see
Fig. \ref{Count1_obs}). We have tested the isolation degree of the
isolated galaxies at $z=1$ also at lower redshifts
($z=0.1,0.3,0.7$). We find that the vast majority of galaxies having
been isolated at $z=1$ or $z=0.5$ were also isolated at these
redshifts. Considering only the subsample of present-day elliptical
isolated galaxies ($B/T > 0.8$), we find that at $z\sim 0.5$ a smaller
fraction of progenitors of present-day early-type galaxies ($\sim 45$
per cent in the Iso\_NN+Q, and $\sim 75$ per cent in the
Iso\_3D$_{\mathrm{RS}}$ sample) have been isolated. At higher
redshift, there is no significant difference between the isolation 
degree of progenitors of isolated early-types and all isolated
galaxies. This indicates that isolated ellipticals have likely
experienced their merger events at $z < 0.5$. We will discuss this in 
more detail in the following section. 

In conclusion, following the progenitors of 2D isolated model
galaxies, we find that a slightly larger fraction of them can be
classified as isolated (or had only few neighbours) in the past (at
$z=0.5,1$) than today. However, we note that we are using co-moving
spheres of 1~Mpc, which corresponds to spheres of $\sim 660$ and $\sim 
500$~kpc in physical units at $z=0.5$ and $z=1$, respectively, what
may explain the slightly higher fractions of isolated progenitors in the
past. As a further consequence, a fixed number of neighbours also
corresponds to a higher probability of interactions at higher redshift
(where galaxies are on average closer). The bottom panel of
Fig.~\ref{Isoevol_obs} shows that at $z=1$, only a small fraction of
$\sim 7$ per cent of the progenitors of 2D isolated galaxies have had
more than 3 neighbours within a co-moving radius of 1~Mpc, which
corresponds to a poor galaxy cluster. 

%*******************************************************************************
\subsection{Minor and major mergers of isolated galaxies}\label{isomerger}
%*******************************************************************************

In order to investigate in more detail the past degree of isolation
for model galaxies selected as isolated at $z=0$, we follow back
in time the main progenitor of each isolated galaxy, and record the
time corresponding to the last merger (only mass-ratios above
1:10). We distinguish between major (mass-ratio of 1:1-1:4) and minor 
mergers (mass-ratio of 1:4-1:10). We find that roughly 45 per cent of
the present-day isolated galaxies (Iso\_NN+Q and
Iso\_3D$_{\mathrm{RS}}$ samples) have had at least one merger event
during their entire life, and one third of these merging galaxies
experienced at least one major merger. This shows that minor mergers
provide a larger contribution to the overall stellar mass  assembly
than major mergers, consistent with earlier studies
(e.g. \citealp{Guo08}). Among the 2D isolated galaxies that
experienced at least one merger, we find that about 24 per cent
experienced more than one mergers, but none had more than
four. I.e. the majority (about 76 per cent) of these galaxies
experienced at most only one merger event during their lifetime, and
in most of the cases this was a minor merger. This result is nearly
independent of the isolated galaxy sample considered (21 per cent of
the 3D isolated galaxies with more than one merger), and it is also
true for the corresponding mass-matched random samples (25 per cent
and 23 per cent for the 2D and 3D mass-matched random samples,
respectively). Therefore on average, isolated galaxies have undergone
\textit{roughly the same amount} of merger events in their life as a 
`typical' galaxy of \textit{similar stellar mass}. This implies that
the isolation degree of our model galaxies of a given stellar mass has
no significant influence on the merging history of their progenitor
galaxies. We stress, however, that merger events that happened at very
early times likely do not have a strong effect on the present-day
properties of galaxies or their isolation degree.

Fig. \ref{LastMerger} illustrates the distribution of the lookback
times since the \textit{last major or minor merger event} (top and
bottom panel, respectively). We find that about 5 per cent of
present-day isolated and mass-matched random galaxies had their last
major or minor merger during the last Gyr. Turning to only major
mergers, the time distributions hardly change for the different galaxy
samples, and since z=3 they are nearly flat for all galaxy samples. On
average, isolated and randomly selected galaxies have had their last
major merger event roughly 6 Gyr ago. In contrast, for minor mergers,
the time distributions are peaking at $z\sim 1$ and the amount of
galaxies having had a minor merger at $z <1 $ decreases with
decreasing time, again independently of the considered galaxy
sample. Our results are consistent with the study by \citet{Guo08},
that is based on the same model used in this study.

Considering the subsample of isolated ellipticals (with $B/T>0.8$),
they exhibit a very different behaviour compared to randomly selected
ellipticals or average isolated galaxies. Fig. \ref{LastMerger_ell}
shows the distributions of the lookback times since the last major or
minor merger event of Iso\_NN+Q and Iso\_3D$_{\mathrm{RS}}$
\textit{elliptical} galaxies and the corresponding mass-matched random
ones. For major mergers (top panel of Fig. \ref{LastMerger_ell}), the
time distributions for isolated ellipticals are skewed towards
decreasing lookback times. Randomly selected ellipticals show a
similar but weaker trend. On average, isolated early-types experienced
their last major merger event roughly 3~Gyr ago, much later than
randomly selected ellipticals (5~Gyr) and than 'average' isolated
galaxies. A similar trend is visible for minor mergers (bottom panel
of Fig. \ref{LastMerger_ell}).  If the last major merger event is
responsible for the formation of the bulge, then our results suggest
that the bulge masses of isolated ellipticals have been assembled on
average only 3~Gyr ago. In contrast, bulges of randomly selected
early-type galaxies of the same mass have been assembled on average
significantly earlier (5~Gyr ago). Note that the recent major merger
is most likely the reason these elliptical galaxies have been
classified as isolated according to the definition adopted. Such a
population would always be expected in a sample of isolated galaxies. 

%*******************************************************************************
%*******************************************************************************
\section{Summary and conclusion}\label{discussion}
%*******************************************************************************
%*******************************************************************************

In this work we have presented a detailed, statistical study of
isolated model galaxies extracted from publicly available galaxy
catalogues based on the merger \textit{trees from} the Millennium
simulation (\citealp{DeLucia07}). We have focused on basic,
present-day properties of model isolated galaxies considering
different selection criteria and based on the same parent galaxy
sample ($m_{\mathrm{b,j}}<15.7$). For our analysis, we have used 2D
criteria widely adopted in the observational literature
(\citealp{Verley07a}, AMIGA sample), as well as a more stringent 3D
selection criterion, where we take advantage of the full 3D-real space
information available in simulations. Our main results can be
summarised as follows:

\begin{itemize}
\item[(i)]{2D selected isolated model galaxies have relatively
    low stellar masses ($<10^{11.5} M_{\odot}$) and a significantly
    high fraction of central galaxies, with respect to galaxies
    residing in high-density regions. Compared to randomly selected 
    samples of model galaxies with the same mass distribution,
    isolated galaxy samples are characterised by a larger fraction of
    late-type star forming galaxies, tend to reside in lower mass
    haloes ($M_{\mathrm{halo}} \leq 10^{13} M_{\odot}$), and contain
    slightly less massive black holes. These trends are in qualitative
    agreement with those of the observed galaxies in the AMIGA sample
    (e.g. \citealp{Verdes05, Sulentic06, Durbala08}) as well as with
    other studies focusing on void galaxies using data from SDSS and
    2dFGS (e.g. \citealp{Hogg03, Hoyle04, Croton05, Patiri06,
      Croton08, Kreckel12, Hoyle12} and references therein). This
    suggests that our results are not specific for the AMIGA sample
    but can/should be considered more general. 

   We find that all 2D samples include a small fraction of satellite
   galaxies, most of which reside (basically by construction) at
   the outskirts of their parent dark matter haloes, where the galaxy 
   number density is lower. 
   Therefore, the criteria that are commonly used to define observed
   isolated galaxy samples, are effective in selecting samples of
   galaxies whose physical properties are in qualitative agreement
   with observational measurements, at least in terms of relative
   trends with respect to the global galaxy population. However, these
   samples are `contaminated' by a fraction of satellite galaxies,
   whose physical properties are influenced by environmental physical
   processes, that ranges between $10$ and $15$ per cent.}  

\item[(ii)]{When comparing different 2D-selection criteria (based on
    cuts in the tidal strength parameter $Q$ and/or in the local
    galaxy number density $\Sigma_{5NN}$), we find that the tidal
    strength parameter represents a stronger constraint for isolation,
    as it excludes small dense systems like pairs or triplets. This
    clearly shows that adding the mass information is important for
    selecting isolated galaxies.  

  Our 3D isolated criterion is based on the same parent sample used
  for the 2D selection, but assumes that galaxies are isolated if they
  have no neighbour within a sphere of 1~Mpc centred on the isolated
  candidate with stellar mass down to one order of magnitude below
  that of the isolated candidate. Adopting these (somewhat arbitrary)
  definitions, we find that roughly two thirds of the 2D isolated
  galaxies are also completely isolated when using the 3D criterion,
  and that most of the `misclassified' 2D isolated model galaxies are
  satellite galaxies. We also show that the 2D sample is `incomplete'
  as roughly one third of 3D  selected isolated galaxies are not
  included in the 2D sample.  In particular, we find that our 2D
  selected samples are significantly `contaminated' at a level of
  about $18$ per cent, and `complete' at a level of $65$ per cent.}   

\item[(iii)]{Our 2D isolated samples contain a very small fraction
    (about $3-4$ per cent) of bulge dominated model galaxies
    (bulge-to-total ratio of $B/T > 0.8$). We find that the bulges of
    these galaxies have been build mainly through merger events, two 
    thirds of which are classified as minor mergers (mass ratio
    between 1:4 and 1:10). The mergers determining the assembly of
    these bulges take place relatively late (about 3 Gyr ago), later
    than the last major merger of a `typical' elliptical galaxy that
    is not classified as isolated (about 5 Gyr ago). 

  Independently of the isolated galaxy sample, about 45 per cent of
  the model galaxies experienced at least one merger event but
  only about 24 per cent of them have had more than one merger events
  and none experienced more than four mergers. Most of the mergers
  (about two thirds) are minor. Almost the same fractions are found
  for a randomly selected sample of galaxies that has the same mass
  distribution of the isolated samples. Therefore, the `isolation' of
  galaxies of given stellar mass does not influence significantly
  their merger history. Thus, we want to point out that the
  differences in the galaxy properties between the isolated and the
  corresponding mass-matched random samples are hardly caused by a
  difference in their merger histories, but mainly by the larger
  fraction of satellites in the random samples, as satellites can
  experience environmental processes.} 

 \item[(iv)] {Using the merger trees available from the simulated
     catalogues, we have studied the degree of isolation of our 2D
     isolated galaxies in the past. Our results show that progenitors
     of isolated galaxies have preferentially resided in `under-dense'
     regions, compared to those of random galaxies. Nevertheless, a
     significant fraction of the galaxies classified as isolated at
     $z=0$ (35-40 per cent for the 2D samples) have not been`isolated'
     since $z=1$ when considering our 3D-real space
     criterion. I.e. following the main progenitors of 2D isolated
     galaxies, we find that a significant fraction of them have
     neighbours within a  comoving sphere of 1~Mpc radius, but only
     7~per cent have more than 3 neighbours at $z=1$. We note that
     this radius corresponds to an increasingly smaller physical
     radius at higher redshift. So a fixed number of neighbours
     corresponds to an increasing probability of interactions at
     increasing redshift. However, we stress that the majority of the
     galaxies classified as isolated (about 88 per cent) have
     experienced at most one minor merger event during their
     life-time, and this merger occurred on average when the Universe
     was half its present age. Therefore, nurture does not play a
     relevant role in the evolution of these galaxies, and the
     approximation that these galaxies have experienced only internal
     physical processes appears to be valid.} 
\end{itemize}

\section*{Acknowledgments}

MH and GDL acknowledge financial support from the European Research
Council under the European Community's Seventh Framework Programme
(FP7/2007-2013)/ERC grant agreement n. 202781. We thank the referee,
Darren Croton, for a careful and constructive reading of our paper.

\bibliographystyle{mn2e}
\bibliography{Literaturdatenbank}

\label{lastpage}

\end{document}